\documentclass[12pt]{article}
\usepackage{amssymb,amsfonts}
\usepackage{graphicx}
\usepackage[top=2.75cm, bottom=2.75cm, left=2.75cm, right=2.75cm]{geometry} 
\usepackage{appendix}
\usepackage{bbm}

\newcommand{\be}{\begin{equation}}
\newcommand{\ee}{\end{equation}}
\newcommand{\beqs}{\begin{eqnarray}}
\newcommand{\eeqs}{\end{eqnarray}}

\def\N{ {\cal N}}
\def\Lag{ {\cal L}}
\def\gcs{{\cal J}}
\def\calT{ {\cal T}}
\def\calI{{\cal I}}
\def\Z{ {\cal Z}}
\def\+{{+\!\!\!+}} 
\def\pp{\mbox{\tiny${}_{\stackrel\+ =}$}} 
\def\vf{{\varphi}}

\newcommand{\pa}{\partial}
\newcommand{\na}{\nabla}
\newcommand{\half}{{\textstyle{\frac12}}}
\newcommand{\Ka}{{K\"ahler}}
\newcommand{\bbX}[1]{\mathbb{X}_{#1}}
\newcommand{\bbXB}[1]{\bar{\mathbb{X}}_{#1}}
\newcommand{\bbD}[1]{\mathbb{D}_{#1}}
\newcommand{\bbDB}[1]{\bar{\mathbb{D}}_{#1}}
\newcommand{\hcd}[1]{\nabla_{#1}}
\newcommand{\cpl}{$\mathbb{P}^1$ }
\newcommand{\z}{\zeta}
\newcommand{\Y}{\Upsilon}
\newcommand{\bY}{{\bar\Upsilon}}
\newcommand{\OO}{ {\cal O}}

\begin{document}
\renewcommand{\theequation}{\thesection.\arabic{equation}}
\begin{titlepage}
\begin{flushright} \small
UUITP-15/12 \\
\end{flushright}
\smallskip
\begin{center} 
\LARGE
{\bf  Supersymmetric Nonlinear Sigma Model Geometry }\normalsize\footnote{Based on  lecures given at ESI, Wienna; Odense; UWA, Perth; Corfu;..}
 \\[12mm] 
{\bf Ulf~Lindstr\"om$^{a}$,} \\[8mm]
{ \small\it
$^a$Department of  Physics and Astronomy, Division of Theoretical Physics,
Uppsala University, \\ Box 516, SE-751 20 Uppsala, Sweden \\}
\end{center}

\vspace{10mm}
\centerline{\bfseries Abstract} 
\bigskip

\noindent  
This is a  review of how sigma models formulated in  Superspace have become important tools
for understanding geometry. Topics included are: The (hyper)k\"ahler reduction; projective superspace; the generalized Legendre construction; generalized K\"ahler geometry and constructions of hyperk\"ahler metrics on Hermitean symmetric spaces.
\vfill
\end{titlepage}
\newpage
\setcounter{page}{1}
\normalsize
\tableofcontents

\section{Introduction}
\label{intro}
Sigma models take their name from a phenomenological model of beta decay introduced more than fifty years ago by Gell-Mann and L\'evy \cite{GellMann:1960np}. It contains  pions and a new scalar meson that they called sigma. A generalization of this model  is what is nowadays meant by a sigma model. It will be described in detail below.

Non-linear sigma models arise in a surprising number of different contexts. Examples are effective field theories (coupled to gauge fields), the scalar sector of supergravity theories, etc. Not the least concern to modern high energy theory is the fact that the string action has the form of a sigma model coupled to twodimensional gravity and that the compactified dimensions carry the target space geometry dictated by supersymmetry.

The close relation between supersymmetric sigma models and complex geometry was 
first observed more thant thirty years ago in \cite{Zumino:1979et} where the target space of $\N=1$ models in four dimensions is shown to carry K\"ahler geometry.  For $\N=2$ models in four dimensions the target space geometry was subsequently shown to be hyperk\"ahler in \cite{Alvarez-Gaume:1981hm}.  This latter fact was extensively exploited in a $\N=1$ superspace formulation of these models in \cite{Lindstrom:1983rt}, where two new constructions were presented; the Legendre transform construction and the hyperk\"ahler quotient construction. The latter reduction was developed and given a more mathematically stringent formulation in \cite{Hitchin:1986ea} where we also elaborated on a manifest $\N=2$ formulation, originally introduced in \cite{Karlhede:1984vr} based on observations in \cite{Gates:1984nk}. 

A $\N=2$ superspace formulation of the $\N=2,$ four dimensional sigma model is obviously desirable, since it will automatically lead to hyperk\"ahler geometry on the target space. The $\N=2$ Projective Superspace
which makes this possible grew out of the development mentioned last  in the preceding paragraph. Over the years it has been developed and refined in, e.g., \cite{Karlhede:1984vr}-\cite{Gonzalez-Rey:1997db}. In this article we report on some of that development along with some more recent development, such as  projective superspace for supergravity \cite{Kuzenko:2007hu}-\cite{Linch:2012zh}, and applications such as the construction of certain classes of hyperk\"ahler metrics \cite{Arai:2006gg}-\cite{Kuzenko:2008ci}.

The target space geometry depends on the number of supersymmetries as well as on the dimension of the domain. There are a number of features peculiar to sigma models with a two dimensional domain (2d sigma models). Here the target space geometry can be  torsionful and generalizes the \Ka \ and hyperk\"ahler geometries. This has  been exploited to give new and interesting results in generalized geometry \cite{Lindstrom:2004eh}
-\cite{Hull:2012dy}.

All our presentations will concern the classical theory. We shall not discuss the  interesting and important question of quantization.

\section{Sigma Models}
\label{sigma}
\setcounter{equation}{0}
A  non-linear sigma model is a theory of maps from a (super) manifold $\Sigma^{(d,\N)}$ to a target space\footnote{We shall mostly avoid global issues and assume that all of $\cal{T}$ can be covered by such maps in patches.}
${\cal T}$:
\beqs\nonumber\label{1}
&&\Phi : \Sigma^{(d,\N)} \rightarrow {\cal{T}}\\[1mm]
&&\Phi(z)\longmapsto Z\in {\cal{T}}~.
\eeqs
Denoting the coordinates on $\Sigma^{(d,\N)}$ by $z=(\xi,\theta)$, the maps are derived by extremizing an action
\be\label{ac1}
S=\int_\Sigma dz{\cal{L}}(\Phi)+\int_{\pa \Sigma}....
\ee
The actual form of the action depends on the bosonic and fermionic dimensions $d_B, d_F$ of $\Sigma$. We have temporarily included a boundary term, which is sometimes needed for open models to have all the symmetries of the bulk-theory \cite{Albertsson:2001dv}-\cite{Lindstrom:2002mc}, or when there are fields living only on the boundary  coupling to the bulk fields as is the well known case for (stacks of) $D$-branes. For a discussion of the latter in a sigma model context, see  \cite{Howe:2005jz}. 
Expanding the superfield as $\Phi(\xi,\theta)=X+\theta\Psi+...$, where $X(\xi)$ and $\Psi(\xi)$ are bosonic and fermionic fields over the even part of $\Sigma$ (coordinatized by $\xi$), the action (\ref{ac1}) becomes
\be\label{ac2}
S= \mu^{d-2}\int_{\Sigma_B} d\xi\left\{\pa X^\mu G_{\mu\nu}(X)\pa X^\nu+\dots\right\}+\int_{\pa \Sigma}....
\ee
where we $d=d_B$ is the bosonic dimension of $\Sigma$ and we have rescaled the $X$'s to make them dimensionless, thus introducing the mass-scale $\mu$.
\bigskip

Let us make a few general comments about the action (\ref{ac2}), mostly quoted from Hull, \cite{Hull:1986hn}:
\bigskip
\begin{enumerate}

\item   The mass-scale $\mu$ shows that the model typically will be non-renormalizable for $d\geqslant 3$ but renormalizable and classically conformally invariant in $d=2$.
\bigskip

\item  We have not included a potential for $X$ and thus excluded Landau-Ginsburg models.
\smallskip

\item  There is also the possibility to include a Wess-Zumino term. We shall return to this when discussing $d=2$.
\bigskip

\item  From a quantum mechanical point of view it is useful to think of $G_{\mu\nu }(X)$ as an infinite number of coupling constants:
\be
G_{\mu\nu}(X)=G^0_{\mu\nu}+G^1_{\mu\nu,\rho }X^\rho+\dots
\ee

\item  Classically, it is more rewarding to emphasize the geometry and think of $G_{\mu\nu}(X)$ as a metric on the target space ${\cal T}$. This is the aspect we shall be mainly concerned with.
\bigskip

\item  The invariance of the action $S$ under {\em Diff}$({\cal T})$,
\be
X^\mu\to X^{\mu'}(X)~,\quad G_{\mu\nu}(X)\to G_{\mu'\nu'}(X')~,
\ee
(field-redefinitions from the point of view of the field theory on $\Sigma$), implies that the sigma model is defined by an equivalence class of metrics. {\em N.B.} This is not a symmetry of the model since the ``coupling constants'' also transform.  It is an important property, however. Classically it means that the model is extendable beyond a single patch in ${\cal T}$, and quantum mechanically it is needed for the effective action to be well defined.
\end{enumerate}
\bigskip

That the geometry of the target space ${\cal T}$ is inherently related to the sigma model is clear already from the preceding comments. Further, the maps extremizing $S$ satisfy
\be\label{2}
\na_i\pa^iX^\mu :=\pa_iX^\nu\na_\nu \pa^iX^\mu=0~,
\ee
where $\pa_i :=\frac \pa {\pa \xi^i }$, the operator $\na$ is the Levi-Civita connection for $G_{\mu\nu}$ and we have ignored the fermions. This is the pull-back of the covariant Laplacian on ${\cal T}$ to $\Sigma$ and hence the maps are sometimes called harmonic maps.

The geometric structure that has emerged from the bosonic part shows that the target space geometry must be Riemannian (by which we mean that it comes equipped with a metric and corresponding Levi-Civita connection). Further restrictions arise from supersymmetry.

\section{Supersymmetry}
\label{Sussi}
\setcounter{equation}{0}
This section provides a very brief summary of some aspects of supersymmetry. For a thorough introduction the reader should consult a textbook, e.g., \cite{Gates:1983nr}-\cite{Buchbinder:1998qv}.

At the level of algebra, supersymmetry is an extension of the $d$-dimensional Poincar\'{e}-algebra to include anticommuting charges $Q$. The form of the algebra depends on $d$. In $d_B=4$ the additional (anti-)commutators satisfy
\be\label{qalg1}
\{Q^a_\alpha,Q^b_\beta\}=2\delta^{ab}(\gamma^iC)_{\alpha\beta}P_i+C_{\alpha\beta}Z^{ab}+(\gamma_5C)_{\alpha\beta}Y^{ab}
\ee
Here $\alpha,\beta,..$ are four dimensional spinor indices, $\gamma^i$ are Dirac gamma matrices, $C$ is the (electric) charge conjugation matrix and the charges $Q^a$ are spinors that satisfy a Majorana reality condition and transform under some internal symmetry group 
${\cal G }\subset O(N)$ (corresponding to the index $a$). The generators of the graded Poincar\'{e}-algebra are thus the Lorentz-generators $M$, the  generators of translations $P$ and the supersymmetry generators 
$Q$. In addition, for non-trivial ${\cal G }$, there are central charge generators $Z$ and $Y$ that commute with all the others.

Representations of supersymmetry are most economically collected into superfields $\Phi(\xi,\theta)$, where $\theta^a$ are Grassmann valued spinorial ``coordinates'' to which one can attach various amounts of importance.
We may think of them as a book-keeping device, much as collecting components into a column-vector.
But thinking about $(\xi,\theta)$ as coordinates on a supermanifold ${\cal M}^{(d,{\cal N})} $ \cite{Salam:1974yz}\cite{BL}  and investigating the geometry of this space has proven a very fruitful way of generating interesting results. 

The index $a$ on $\theta^a$ is the same as that on $Q^a$ and thus corresponds to the number ${\cal N}$ of supersymmetries. Let us consider the case ${\cal N}=1,~d_B=4$, which implies $Z=Y=0$. In a Weyl-representation
of the spinors and with the usual identification of the translation generator as a differential operator
$P_{\alpha\dot\alpha} =i\pa_{\alpha\dot \alpha}:=i\sigma_{\alpha\dot \alpha}^i\pa_i$, the algebra (\ref{qalg1}) becomes
\be\label{apmage}
\{Q_\alpha,\bar Q_{\dot \alpha}\}=2i\pa_{\alpha\dot \alpha}
\ee
Introducing Berezin integration/derivation \cite{Berezin:1966nc}, $\pa_\alpha:=\frac \pa {\pa \theta^\alpha }$,  the supercharges $Q$ may also be represented (in one of several possible representations) as differential operators acting on superfields:
\beqs\label{33}
&& Q_\alpha=i\pa_\alpha+\frac 1 2 \bar \theta^{\dot \alpha}\pa_{\alpha\dot \alpha}\cr
&&\bar Q_{\dot \alpha}=i\pa_{\dot \alpha}+\frac 1 2  \theta^{\alpha}\pa_{\alpha\dot \alpha}~.
\eeqs
Apart from the various representations alluded to above (chiral, antichiral and vector in four dimensions \cite{Gates:1983nr}) there are  two basic ways that  supercharges can act on superfields, corresponding to left and right group action. This means that, given  (\ref{33}), there is a second pair of differential operators
\beqs
&& D_\alpha=\pa_\alpha+i\frac 1 2 \bar \theta^{\dot \alpha}\pa_{\alpha\dot \alpha}\cr
&&\bar D_{\dot \alpha}=\pa_{\dot \alpha}+i\frac 1 2  \theta^{\alpha}\pa_{\alpha\dot \alpha}~,
\eeqs
which also generate the algebra (\ref{apmage}) and that anticommute with the $Q$'s:
\be\label{ardvag}
\{D_\alpha,\bar D_{\dot \alpha}\}=2i\pa_{\alpha\dot \alpha}~,\quad \{Q_\alpha,\bar D_{\dot \alpha}\}=\{Q_\alpha,D_{\beta}\}=0~.
\ee
Often the supersymmetry algebra is given only in terms of the $D$'s.
From a geometrical point of view, these are covariant derivatives in superspace and may be used to impose invariant conditions on superfields. 

In general, covariant derivatives $\na_{\cal A}$ in a curved superspace space satisfy
\be\label{curve}
[\na_{\cal A},\na_{\cal B}\}=R_{\cal A\cal B}\cdot\mathbb{M}+2T_{\cal A\cal B}^{~~~\cal C}\na_{\cal C}~,
\ee
where the left hand side contains a graded commutator $T_{\cal A\cal B}^{~~~\cal C}$ is the torsion tensor and 
$R_{\cal A\cal B}\cdot\mathbb{M}$ the curvature  with $\mathbb{M}$ the generators of the structure group. The indices ${\cal A}$ etc run over both bosonic and fermionic indices. Comparision  to (\ref{ardvag}), with $\na_{\cal A}=(D_\alpha,\bar D_{\dot \alpha},i\na_{\alpha\dot\alpha})$, shows that even ``flat'' ($R_{\cal A\cal B}\cdot\mathbb{M}=0$) superspace has torsion.

In four dimensions the $D$'s may be used to find the smallest superfield representation. Such a chiral superfield $\phi$ and its complex conjugate antichiral field $\bar\phi$
are required to satisfy
\be\label{chiral}
\bar D_{\dot \alpha}\phi=D_\alpha \bar\phi=0~.
\ee
The Minkowski-field content of this may be read off from the $\theta$-expansion. However, this expansion depends on the particular representation of the $D$'s and $Q$'.  For this reason it is preferable to define the components in the following representation independent form:
\be\label{thzero}
X:=\phi|~,~~~\chi:=\bar D\phi|~,~~~\bar \chi:=D\phi|~,~~~{\cal F}:=\half D\bar D \phi|~,
\ee
where a vertical bar denotes ``the $\theta$-independent part of''. In a chiral representation where $\bar D_\alpha=\partial_\alpha$ the $\theta$-expansion of a chiral field reads
\be
\phi(\xi,\theta)=X(\xi)+\bar\theta\chi(\xi)+\theta\bar\chi(\xi)+\bar\theta\theta{\cal F}~,
\ee
but its complex conjugate involves a $\theta$ dependent shift in $\xi$ and looks more complicated.

A dimensional analysis shows that if $X$ is a physical scalar, $\chi$ is a physical spinor and ${\cal F}$ has to be a non-propagating (auxiliary) field. This is the smallest multiplet that contains a scalar, and thus suitable for constructing a supersymmetric extension of the bosonic sigma models we looked at so far\footnote{All other superfields will either be equivalent to (anti) chiral ones or contain additional bosonic fields of higher spin.}. Denoting a collection of chiral fields by $\phi =(\phi^\mu)$, the most general action we can write down
\be\label{serres}
S=\int d^4\xi d^2\theta d^2\bar \theta K(\phi,\bar \phi)
\ee
reduces to the bosonic integral
\be
\int d^4\xi\left\{\pa^iX^a\pa_b\pa_{\bar a}K(X,\bar X)\pa_i\bar X^{\bar a}+\dots\right\}~.
\ee
The most direct way to perform the reduction is to write
\be\label{xanthi}
S=\int d^4\xi D^2\bar D^2K(\phi,\bar \phi)|~,
\ee
and then to use (\ref{chiral}) when acting with the covariant spinor derivatives.
Note that $K$ in (\ref{serres})  is only defined up to a term $\eta(\phi)+\bar\eta(\bar\phi)$ due to the chirality conditions (\ref{chiral}).

We immediately learn additional things about the geometry of the target space ${\cal T}$:
\begin{enumerate}
\item  It must be even-dimensional. 
\item  The metric is hermitean with respect to the canonical complex structure 
\be
J:=\left( \begin{array}{cc}
i\delta^a_b&0\\
0&-i\delta^{\bar a}_{\bar b}\end{array}\right)
\ee
for which $X$ and $\bar X$ are canonical coordinates. 
\item  The metric has a potential $K(\phi,\bar\phi)$. In fact, the geometry is K\"ahler and the ambiguity in the Lagrangian in (\ref{serres}) is known as a \Ka \ gauge transformation.
\end{enumerate}

\section{Complex Geometry I}
\setcounter{equation}{0}

Let us interrupt the description of sigma models to recapitulate the essentials of K\"ahler geometry.

Consider $({\cal M},G,J)$ where G is a metric on the manifold $\cal M$ and $J$ is an almost complex structure, i.e., an endomorphism\footnote{A $(1,1)$ tensor  $J^\mu_{~\nu}$.} $J: TM\hookleftarrow $ such that $J^2=-1$. (Only possible if $\cal M$ is even dimensional.) This is an almost hermitean space if  (as matrices)
\be
J^tGJ=G~.
\ee
Construct the projection operators
\be
\pi_{\pm}:=\half (1\pm iJ)~.
\ee
If the vectors $\pi_\pm V$ in $T$  are in involution, i.e., if
\be\label{invol}
\pi_\mp[\pi_\pm V,\pi_\pm U]=0~,
\ee
which implies that the Nijenhuis torsion vanishes,
\beqs\label{Nij}
N_J(U,V)=[JU,JV]-J[JU,V]-J[U,JV]-[U,V]=0~,
\eeqs
then the distributions defined by $\pi_\pm$ are integrable and $J$ is called a complex structure and $G$  hermitean.
To evaluate the expression in (\ref{invol}) in index notation, think of $\pi_\mp$ as matrices acting on the components of the Lie bracket  between the vectors 
$\pi_\pm V$ and $\pi_\pm U$. This leads to the following expression corresponding to (\ref{Nij}):
 \be
\mathcal{N}(J)^\mu_{\nu\rho} = J^\sigma_{[\nu}J^\mu_{\rho],\sigma}+J^\mu_\sigma J^\sigma_{[\nu,\rho]}=0~.
\ee

The fundamental two-form $\omega$ defined by $J$ and $G$ is
\be\label{holtwo}
\omega:=G_{\mu\nu}J^\nu_{~\sigma}dX^\mu\wedge dX^\sigma~.
\ee
If it is closed for a hermitean complex space, then the metric has a K\"ahler potential
\be
G=\left(\begin{array}{cc}
0&\pa\bar\pa K\\
\bar\pa\pa K&0\end{array}\right)
\ee
and the geometry is K\"ahler. 

An equivalent  characterization is as an almost hermitean manifold with
\be
\na J=0~,
\ee 
where $\na$ is the Levi-Civita connection.

We shall also need the notion of hyperk\"ahler geometry. Briefly, for such a geometry there exists an
$SU(2)$-worth of complex structures labeled by $A\in \{1,2,3\}$
\be
J^{(A)}J^{(B)}=-\delta^{AB}+\epsilon^{ABC}J^{(C)}~, \quad \na J^{(A)}=0, ~~~
\ee
with respect to all of which the metric is hermitean
\be
J^{(A)t}GJ^{(A)}=G~.
\ee

\section{Sigma model geometry}
\setcounter{equation}{0}

We have already seen how ${\cal N}=1$ (one supersymmetry) in $d=4$ requires the target space geometry to be  K\"ahler. To investigate how the geometry gets further restricted for ${\cal N}=2$, there are various options: i) Discuss the problem entirely in components (no supersymmetry manifest). ii) Introduce projective (or harmonic) superspace and write models with manifest ${\cal N}=2$ symmetry. iii) Add a non-manifest supersymmetry to the model already descibed and work out the consequences.
The last option requires the least new machinery, so we first follow this.
The question is thus under what conditions
\be
S=\int d^4\xi d^2\theta d^2\bar \theta K(\phi,\bar \phi)
\ee
can support an additional supersymmetry. The most general ansats for such a symmetry is
\be
\delta\phi^\mu =\bar D^2(\bar\varepsilon\bar\Omega^\mu)~.
\ee
One finds closure of the additional supersymmetry algebra (on-shell) and invariance of the action provided that the target space geometry is hyperk\"ahler with the non-manifest complex structures formed from 
$\Omega$ \cite{Hull:1985pq}:
\be
J^{(1)}=\left(\begin{array}{cc}
0&\Omega^{\bar a}_{~,b}\\
\bar\Omega^a_{~,\bar b} &0\end{array}\right)\quad 
J^{(2)}=\left(\begin{array}{cc}
0&i\Omega^{\bar a}_{~,b}\\
-i\bar\Omega^a_{~,\bar b} &0\end{array}\right)\quad
J^{(3)}=\left(\begin{array}{cc}
i{\mathbbm 1}&0\\
0 &-i{\mathbbm 1}\end{array}\right)
\ee
Further, the parameter superfield $\varepsilon$ obeys
\be
\bar D_{\dot \alpha}\varepsilon=D^2 \varepsilon=\partial_{\alpha \dot\alpha}\varepsilon =0~.
\ee
It contains the parameter for central charge transformations along with the supersymmetry parameters.\\
The full table of geometries for supersymmetric non-linear sigma models without Wess-Zumino term reads
\bigskip
  \begin{center}
  \begin{tabular}[htb]{|l|lll|l| }
  \hline
  d=&6&4&2&Geometry\cr
  \hline
  {\cal N}=&1&2&4&Hyperk\"ahler\cr
  {\cal N}=&~&1&2&K\"ahler\cr
  {\cal N}=&~&~&1&Riemannian\cr
  \hline
  \end{tabular}
  \end{center}
  \bigskip
(Odd dimensions have the same structure as the even dimension lower.) When we specialize to two or six dimensions, we have the additional possibility of having independent left and right supersymmetries; 
the ${\cal N}=(p,q)$ supersymmetries of Hull and Witten \cite{Hull:1985jv}. We shall return to this possibility when we discuss $d=2$, but now we turn to the question of how to gauge isometries on K\"ahler and hyperk\"ahler manifolds.

\eject
\section{Gauging isometries and the HK reduction}
\label{HK}
This section is to a large extent a review of  \cite{Hull:1985pq} and \cite{Hitchin:1986ea}.

\subsection{Gauging isometries of bosonic sigma models}
\setcounter{equation}{0}

The table in the previous section describing the target-space geometry shows that constructing, e.g., new $ {\cal N}=2, d=4$ nonlinear sigma models is tantamount to finding new hyper-k\"ahler geometries. A systematic method for doing this involves isometries of the target space, which we now discuss.

Consider again the bosonic action 
\be\label{acta}
S=\int d\xi \partial_i \phi^\mu G_{\mu\nu}(\phi)\partial^i\phi^\nu~.
\ee
As noted in Sec.\ref{sigma}, a target space diffeomorphism leaves this action invariant and corresponds to a field redefinition. As also pointed out, this is not a symmetry of the field theory. A symmetry of the field theory involves a transformation of $\phi$ only;
\be
\delta\phi^\mu = \lambda^Ak^\mu_A(\phi)=[\lambda k,\phi]^\mu \equiv{\cal{L}}_{\lambda k}\phi^\mu~,
\ee
where ${\cal{L}}_{\lambda k}$ denotes the Lie derivative along the vector $\lambda k$.
Under such a transformation the action varies as
\be
\delta S=\int d\xi \partial_i \phi^\mu {\cal{L}}_{\lambda k} G_{\mu\nu}(\phi)\partial^i\phi^\nu ~.
\ee
The transformation thus gives an invariance of the action if
\be
{\cal{L}}_{\lambda k} G_{\mu\nu}=0~,
\ee
i.e., if the transformation is an isometry and hence  if the $k^\mu_A$'s are Killing-vectors.

We take the $k_A$'s to generate a Lie algebra ${\frak g}$
\be\label{algb}
[k_A,k_B]=c_{AB}^{~~~C}k_C
\ee with 
\be
k_A\equiv k^\mu_A\frac{\partial}{\partial \phi^\mu}~.
\ee
In what follows, we assume that  ${\frak g}$ can be exponentiated to a group $\cal G$.

One way to construct a new sigma model  from one which has isometries is to gauge the isometries and then find a gauge connection that extremizes the action
\cite{Hull:1985pq}. The new sigma model will be a quotient of the original one. Briefly, this goes as follows:

The isometries generated by $k_A$ are gauged introducing a gauge field $A^A_i$ using minimal coupling,
\be
\partial_i\phi^\mu\to\partial_i\phi^\mu-A^A_ik^\mu_A=(\partial_i-A^A_ik_A)\phi^\mu\equiv \na_i\phi^\mu~,
\ee
in the action (\ref{acta}):
\be\label{actad}
S=\int d\xi \na_i \phi^\mu G_{\mu\nu}(\phi)\na^i\phi^\nu~.
\ee
This action is now locally invariant under the symmetries defined by the algebra (\ref{algb}). 
Note that there is no kinetic term for the gauge-field.
Extremizing (\ref{actad}) with respect to $A^A_i$ singles out a particular gauge-field:
\be
\delta S=0 \Rightarrow A^A_i=\mathbb{H}^{-1AB}G_{\mu\nu}k^\mu_B\partial_i\phi^\nu
\ee
where
\be
\mathbb{H}_{AB}\equiv k^\mu_AG_{\mu\nu}k^\nu_B~.
\ee
In terms of this particular connection, the action (\ref{actad}) now reads
\be\label{actare}
S=\int d\xi \partial_i \phi^\mu\left( G_{\mu\nu}
-\mathbb{H}^{-1AB}k_{\mu_A}k_{\nu_B}\right)\partial^i\phi^\nu~,
\ee
where indices have been lowered using the metric $G$. Since 
$ \partial_i \phi^\mu\approx \partial_i \phi^\mu+v^{A}_{i}k_{A}^\mu$ in the action (\ref{actare} ), this is a new
 sigma model defined on the 
space of orbits of the group $\cal G$, i.e., on the quotient space $\cal T/\cal G$. The metric on this space is
$\tilde G_{\mu\nu}\equiv G_{\mu\nu}
-\mathbb{H}^{-1AB}k_{\mu_A}k_{\nu_B}$.

To apply this construction to K\"ahler manifolds, or equivalently, supersymmetric models, in such a way as to preserve the 
K\"ahler properties more restrictions are required. First, the isometries we need to gauge are holomorphic and the gauge group
we have to consider is the complexification of the group relevant to the bosonic part.

\subsection{Holomorphic isometries}

A holomorphic isometry on a K\"ahler manifold satisfies
\be
\Lag_{\lambda k}J=0, \quad \Lag_{\lambda k}\omega=0~,
\ee
where $J$ is the complex structure and $\omega$ the K\"ahler two-form defined in (\ref{holtwo}). 
The fact that a K\"ahler manifold is symplectic makes it possible to consider the moment map for 
the Hamiltonian vector field $\lambda k$. The corresponding Hamiltonian function $\mu^{\lambda k}$ 
is defined by
\be
\imath _{\lambda k}\omega=:d\mu^{\lambda k}~
\ee
In holomorphic coordinates $(\phi^{q},\bar\phi^{\bar q})$ where $\lambda k=\lambda^{A}(k_{A}+\bar k_{A})$, this reads
\be
\omega_{\bar q p}\lambda^{A}\bar k_{A}^{\bar q}\equiv-iK_{\bar q p}\lambda^{A}\bar k_{A}^{\bar q}=\frac{\partial \mu^{\lambda k}}{\partial\phi^p}=:\mu^{\lambda k}_{p}~.
\ee
From these relations it is clear why $\mu^{\lambda k}$  is sometimes referred to as a Killing potential.
Now $\mu$ defines a map from the target space of the sigma model into the dual of the Lie-algebra generated by $k_{A}$:
\be
{\cal{T}} \mapsto {}^*\mathfrak{g}, \quad \mu^{\lambda k}=\lambda^{A}\mu_{A}~,
\ee
where $\mu_{A}$ is the basis for  ${}^*\mathfrak{g}$ that corresponds to the basis $k_{A}$ for $\mathfrak{g}$.
When the action of the Hamiltonian field can be made to agree witht the natural action of the group $\cal{G}$ on $\cal{T}$ and on 
${}^*\mathfrak{g}$, the $\mu_{A}$'s are called moment maps\footnote{We use to the (US) East cost nomenclature as opposed to the West coast ``momentum map''.}. This is the case when $\mu^{\lambda k}$ is equivariant, i.e., when
\beqs
&&\lambda k\mu^{\lambda' k}=\mu^{[\lambda k,\lambda' k]}\cr
\iff && k_{A}^p\mu_{B,p}+\bar k_{A}^{\bar p}\mu_{B,\bar p}=c_{AB}^{~~~C}\mu_{C}~,
\eeqs
where the equivalence refers to holomorphic isometries.

The K\"ahler metric is the Hessian of the K\"ahler potential  $K$. As mentioned in Sec.\ref{Sussi}, this leaves an ambiguity in the potential; it is only defined up to the 
sum of a holomorphic and an antiholomorphic term.
\be
K(\phi,\bar \phi) \approx K(\phi,\bar \phi) + \eta(\phi) + \bar\eta(\bar \phi)~.
\ee
An isometry thus only has to preserve $K$ up to such terms:
\be\label{iskogtf}
\Lag_{\lambda k}K=\eta ^{\lambda k}+ \bar\eta^{\lambda k}~.
\ee
For a holomorphic isometry $\lambda k=\lambda^{A}(k_{A}(\phi)+\bar k_{A}(\bar \phi))$ this implies for the projection
\be\label{momrel}
\half \left(1-iJ\right)\lambda k K=\lambda^{A}k_{A}(\phi)\partial_{p}K=-i\mu^{\lambda k}+\eta^{\lambda k}~.
\ee
(Recall that $\lambda ^{A}\bar k_{A}^{\bar q}K_{p\bar q}=i\lambda^{A}\mu_{A,p}$ and 
$\lambda ^{A} k_{A}^{q}K_{q\bar p}=-i\lambda^{A}\mu_{A,\bar p}$, so that 
$\lambda ^{A} k_{A}^{q}K_{q}=-i\lambda^{A}\mu_{A}$+hol. and
 $\lambda ^{A}\bar k_{A}^{\bar q}K_{\bar q}=i\lambda^{A}\mu_{A}$+antihol. )
 
 \subsection{Gauging isometries of supersymmetric sigma models}
 
 Due to the chiral nature of superspace, the isometries act through the complexification of the isometry group $\cal{G}$.
 Explicitly, the parameter $\lambda$ gets replaced by superfield parameters $\Lambda$ and  $\bar\Lambda$, while the bosonic 
 gauge field $A^{A}_i$ becomes one of the components of a real superfield $V^{A}$. (This is the last occurance of $i$ as a world volume index.  Below  $i,j,...$ denote gauge indies.)
 In the simplest case of  isotropy,
 $k^i_A(\phi)= (T_A)^i_j\phi^j$, the coupling to the chiral fields in the sigma model is 
 \be\label{tilde}
 \bar \phi^{i} \to \tilde \phi^i := \bar \phi^j(e^V)_{j}^{~i}
~,\qquad (V)_{j}^{~i}=V^A(T_A)_{j}^{~i}~.
 \ee
 At the same time, a gauge transformation with (chiral) parameter $\Lambda$ acts on the chiral and antichiral fields as
 \be
\bar \phi^{i} \to \bar \phi^j(e^{-i\bar\Lambda})_{j}^{~i}~,\qquad (\bar\Lambda)_{j}^{~i}=\bar\Lambda^A(T_A)_{j}^{~i}~.
 \ee
 The relation (\ref{tilde}) can thus be interpreted as a gauge transformation of $\bar \phi$ with parameter $iV$. For general isometries a gauge transformation with parameter $iV$ is;
 \be
 \bar \phi^{i} \to \tilde \phi^i \equiv e^{\Lag_{iVk}}\bar \phi~,
 \ee
 and this is the form we need to use when defining $\tilde \phi ^i$.
 The coupling is thus
 \be\label{holgauge}
 K(\phi,\bar\phi)\to K(\phi,\tilde\phi)~.
 \ee
 Under a global isometry transformation, according to (\ref{iskogtf}), there may arise terms such as
 \be
 \delta S=\int \lambda^{A}\bar\eta_{A}(\bar\phi) = 0
 \ee
whose vanishing ensures the invariance.
 This is no-longer true in the local case where the corresponding term
 \be
 \delta S=\int \Lambda^{A}\tilde\eta_{A}
 \ee
 will not vanish in general. The remedy is to introduce auxiliary coordinates $(\zeta, \bar \zeta)$ and assign  transformations to them
that make the modified K\"ahler potential $\tilde K$ invariant (rather than invariant up to (anti) holomorphic terms):
\be
\tilde K:=K(\phi, \bar \phi)-\zeta-\bar \zeta~.
\ee
The action involving  $\tilde K$ may now be gauged using the prescription (\ref{holgauge}) and takes the form (dropping the irrelevant $\zeta+\bar\zeta$ term after gauging)
\be
\tilde S=\int \hat K(\phi, \bar \phi, V) ~,
\ee
where 
\be\label{hatK}
\hat K\equiv K(\phi,\tilde \phi)-i\frac{e^{\Lag_{iVk}}-1}{\Lag_{iVk}}\bar\eta_{A}V^{A}~.
\ee
Using the definition of $\tilde \phi$ and the relation to the moment maps (see \ref{momrel} and below),
we rewrite this as
\be
\hat K= K(\phi,\bar\phi)+\left(\frac{e^{\Lag_{iVk}}-1}{\Lag_{iVk}}\right)V^{A}\mu_{A}~.
\ee
A more geometric form of the gauged Lagrangian is
\be\label{geomlag}
\hat K=K(\phi,\bar\phi)+\int_{0}^1dt e^{-\half t\Lag_{VJk}}\mu^V~,
\ee
where we recall that
\beqs
&&\Lag_{k}K=\eta+\bar \eta\cr
&&\Lag_{Jk}K=i(\eta-\bar \eta)+2\mu~.
\eeqs
The form of the action that follows from (\ref{geomlag}) directly leads to the symplectic quotient as applied to a K\"ahler manifold \cite{stern}: Eliminating $V^{A}$ results in
\beqs\label{sympq}
&&e^{-\half t\Lag_{VJk}}\mu_{A}=0, \cr
\iff &&  \mu_{A}=0~.
\eeqs
 The K\"ahler quotient is illustrated in the following picture, taken from \cite{Hitchin:1986ea}:
 
\includegraphics[scale=0.55]{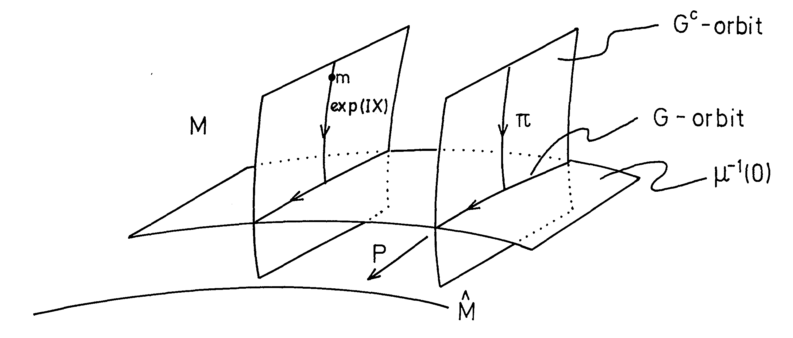}\\
The isometry group $\cal G$ acts on $\mu^{-1}(0)$ and produces the quotient $\hat M$. The same space is obtained if one considers the extension of 
$\mu^{-1}(0)$ by $exp(JX)$ and takes the quotient by the complexified group $\cal G^C$.
\bigskip

If we start from a hyperk\"ahler manifold with tri-holomorphic isometries, there will also be complex moment maps corresponding to 
the two non-canonical complex structures
\be
\omega^\pm \leftrightarrow \mu^\pm~.
\ee
 In addition to (\ref{sympq}),  we will then  have the conditions
 \be
 \mu^+_{A}=\mu^-_{A}=0~,
 \ee
 defining holomorphic subspaces. This hyperk\"ahler quotient prescription \cite{Lindstrom:1983rt}\cite{Hitchin:1986ea}, gives a new hyperk\"ahler space from an old one.
 The ${\cal{N}}=2$ sigma model action that encodes this (in $ d=4, {\cal{N}}=1$ language) reads
 \be
 S=\int d^4\xi\left(d^2\theta d^2\bar\theta\hat K(\phi, \bar \phi, V)+\half d^2\theta S^{A}\mu^+_{ A}
 +\half d^2\bar\theta \bar S^{A}\mu^-_{ A}\right)~,
 \ee
 where $\hat K$ is defined in (\ref{hatK}) and $(V,S,\bar S)$ is the ${\cal{N}}=2$ vector (gauge) multiplet. The latter consists of a chiral superfield $S$ and its complex conjugate $\bar S$ in addition to $V$.

\section{Two dimensional models and generalized K\"ahler geometry}
\setcounter{equation}{0}

The quotient constructions just discussed are limited to backgrounds with only a metric present. Recently extensions of such geometries to include also an antisymmetric $B$-field
have led to a number of new results in $d=2$ which are expected to contribute to new quotients involving such geometries.

Two (bosonic) dimensional domains $\Sigma$ are interesting in that they support sigma models with independent left and right supersymmetries. Such $(p,q)$ models were introduced by Hull and Witten in \cite{Hull:1985jv}, 
and also have analogues in $d=6$. There is a wealth of results on the target-space geometry for $(p,q)$-models. The geometry is typically a generalization of K\"ahler geometry with vector potential for the 
metric instead of a scalar potential et.c. See, e.g., \cite{Hull:1991uw}\cite{Hull:1993ct}. Here we first focus  on $(1,1)$ and $(2,2)$ models in  $d=2$.

The $(1,1)$ supersymmetry algebra is 
\be
D^2_\pm=i\pa_{\pp}~,
\ee
where $+$ and $-$ are spinor indices and $\xi^\+,\xi^=$ are light-cone coordinates in $d=2$ Minkowski space.

A general sigma model written in terms of real ${\cal N}=(1,1)$ superfields $\Phi$ is
\label{2action}
\be
S=\int_\Sigma d^2\xi d^2\theta D_+\Phi^\mu E_{\mu\nu }(\Phi)D_-\Phi^\nu~,
\ee
where the metric $G$ and $B$-field have been collected into
\be
E_{\mu\nu }\equiv G_{\mu\nu }+B_{\mu\nu }
\ee
Here ${\cal N}=(1,1)$ supersymmetry is manifest by construction and we shall see that additional non-manifest 
ones will again restrict the target space geometry. In fact, the geometry is already modified due to the presence of 
the $B$-field. The field-equations now read
\be
\na^{(+)}_{+}D_{-}\Phi^\mu=0~,
\ee
where 
\be\label{nablator}
\na^{(\pm)}\equiv\na^{(0)}\pm T~,
\ee
is the sum of the Levi-Civita connection and a torsion-term formed from the field-strength for the
$B$-field\footnote{The anti-symmetrization does not include a combinatorial factor.}:
\be
T_{\mu\nu\rho}\equiv H_{\mu\nu\rho}=\half\partial_{[\mu}B_{\nu\rho]}~.
\ee
Only when $H=0$ do we recover the non-torsionful Riemann geometry.
The full picture, is given in the following table:
\bigskip

\begin{table}[htbp]
\centering
\begin{tabular}{|l|c|c|c|c|c|}
\hline
Supersymmetry & (0,0) or (1,1) & (2,2) &{ (2,2)}& (4,4) & (4,4) \\ 
\hline
Background & $G,B$ &  $G$ & ${ G,B}$ & $G$ & $G,B$ \\
\hline
Geometry   &  Riemannian & K\"ahler & { { bihermitian}}  & hyperk\"ahler & bihypercomplex \\
\hline
\end{tabular}
\caption{The geometries of sigma-models with different supersymmetries.}
\end{table}
\bigskip

We first look at the Gates-Hull-Ro\v cek (GHR) bihermitean geometry, or Generalized K\"ahler geometry
as is its modern guise. Starting from the ${\cal N}=(1,1)$ action (\ref{2action}), one can ask for additional,
non-manifest supersymmetries. By  dimensional arguments, such a symmetry must act on the superfields as
\be
\delta^{(\pm)}\Phi^\mu=\epsilon^{\pm}D_{\pm}\Phi^\nu J_{\nu}^{(\pm)\mu}(\Phi)~,
\ee
where ${(\pm)}$ correpsond to left or right symmetries. It was shown by GHR in \cite{Gates:1984nk} that invariance of the action
(\ref{2action}) and closure of the algebra require that $J^{(\pm)}$ are complex structures that are covariantly constant 
with respect to the  torsionful connections
\be
\na^{(\pm)}J^{(\pm)}=0~,
\ee
and that the metric is hermitean with repect to both these complex structures
\be
J^{(\pm)t}GJ^{(\pm)}=G~.
\ee
In addition, the $B$-field field-strength (torsion) must obey
\be
H=H^{(2,1)}+H^{(1,2)}=d^c_{(+)}\omega_{(+)}=-d^c_{(-)}\omega_{(-)}~,
\ee
where the chirality assignment refers to {\em both} complex structures.
Here we have introduced $d^c:= J(d)=i(\bar\partial-\partial)$, where the last equality holds in canonical coordinates,
the two-forms  are $\omega_{(\pm)}\equiv GJ^{(\pm)}$ as tensors and
$dd^c_{(\pm)}\omega_{(\pm)}=0$. When $J^{(+)}=\pm J^{(-)}$ this geometry reduces to K\"ahler geometry.

Gualtieri  gives a nice interpretation of the full bihermitean geometry in the context of Generalized Complex Geometry
\cite{hitchinCY} and calls it Generalized K\"ahler Geometry \cite{gualtieri}, which we now briefly describe.

\section{Complex Geometry II}
\setcounter{equation}{0}

The definition of Generalized Complex Geometry (GCG) paralells that of complex geometry but is based on the sum of the tangent and 
cotangent bundle instead of just the tangent bundle. Hence we consider a section $\gcs$ of $End(\calT \oplus \calT ^*)$ such that
$\gcs^2=-1$. To define integrability we again use projection operators
\be
\Pi_{\pm}\equiv \half\left(1\pm i\gcs\right)~,
\ee
but now we require that the subspaces of $\calT \oplus \calT ^*$ defined by  $\Pi_{\pm}$  are in involution with respect to a 
bracket defined on that bundle. Denoting an element of $\calT \oplus \calT ^*$ by $v+\xi$ with $v\in \calT, \xi \in \calT^*$, 
we thus require
\be
\Pi_{\mp}\left[ \Pi_{\pm}(v+\xi),\Pi_{\pm}(u+\chi)\right]_{C}=0~,
\ee
where the bracket is the Courant bracket defined by
\be
\left[ v+\xi,u+\chi\right]_{C}=\left[v,u\right]+\Lag_{v}\chi-\Lag_{u}\xi-\half d(\imath _{v}\chi-\imath _{u}\xi)~,
\ee
where $[~,~]$ is the Lie bracket and,e.g., $\imath _{v}\chi=\chi(v)=v\cdot\chi$.
The full definition of GCG also requires the natural pairing metric $\calI$ to be preserved. The natural pairing is
\be\label{natpair}
<v+\xi,u+\chi>= \imath _{v}\chi+\imath _{u}\xi~.
\ee
In a coordinate basis $(\pa_{\mu},dx^\nu)$ where  we represent $v+\xi$ as $(v,\xi)^t$ the relation in (\ref{natpair}) may be written as 
\cite{Lindstrom:2004iw}:

\be
\left(v,\xi\right)\calI\left(\begin{array}{c}u\\
\chi \end{array}\right)=\left(\begin{array}{cc}
0&v\chi\\
u\xi &0\end{array}\right)~,
\ee
so that 
\be
\calI=\left(\begin{array}{cc}
0&1\\
1 &0\end{array}\right)~,
\ee
and preservation of $\calI$ by  $\gcs$ means 
\be
\gcs^t\calI\gcs=\calI~.
\ee
The specialization to Generalized K\"ahler Geometry (GKG) occurs when we have {\em two} commuting GCS's 
\be
\left[\gcs^{(1)},\gcs^{(2)}\right]=0~.
\ee
This allows the definition of a metric\footnote{This is really a local product structure as defined, 
but appropriate contractions with $\calI$
makes it a metric.} ${\cal G}\equiv-{\gcs}^{(1)}{\gcs}^{(2)}$ which satisfies 
\be
{\cal G}^2=1~.
\ee
The definition of GKG requires this metric to be positive definite.

The relation of GKG to bihermitean geometry is given by the following ``Gualtieri map'':
\be\label{Gmap}
\gcs^{(1,2)}=\left(\begin{array}{cc}
1&0\cr
B&1\end{array}\right)\left(\begin{array}{cc}
J^{(+)}+\pm J^{(-)}&-(\omega_{(+)}^{-1}\mp\omega_{(-)}^{-1})\cr
\omega_{(+)}\mp\omega_{(-)}&-(J^{t(+)}+\pm J^{t(-)})\end{array}\right)
\left(\begin{array}{cc}
1&0\cr
-B&1\end{array}\right)
\ee
which maps the bihermitean data into the GK data. When $H=0$ the first and last matrix on the right hand side represent a 
``$B$-transform'' which is one of the automorphisms of the Courant bracket, but here $H\ne 0$ in general.

For the K\"ahler case, $\gcs^{(1,2)}$ in (\ref{Gmap}), and ${\cal G}$ reduce to
\beqs
\gcs^{(1)}=\left(\begin{array}{cc}
J&0\cr
0&-J^t\end{array}\right)~,~~~\gcs^{(2)}=\left(\begin{array}{cc}
0&-\omega^{-1}\cr
\omega&0\end{array}\right)~,~~~{\cal G}=\left(\begin{array}{cc}
0&g^{-1}\cr
g&0\end{array}\right)~,
\eeqs
where $J$ is the complex structure, $\omega$ is the corresponding K\"ahler form and $g$ is the hermitean metric. This fact is the origin of the name Generalized K\"ahler coined by Gualtieri \cite{gualtieri}.

\section{${\cal N}=(2,2)~, d=2$ sigma models off-shell}
\label{offshell}
\setcounter{equation}{0}

The $\N=(1,1)$ discussion of GHR identified the geometry of the sigma models that could be extended to have $\N=(2,2)$ supersymmetry,
but they found closure of the algebra only when the complex structures commute, i.e., on  $ker[J^{(+)},J^{(-)}]$, in which case they gave a full
$\N=(2,2)$ description in terms of chiral and twisted chiral fields. In this section we extend the discussion to the non-commuting case and describe 
the general situation following \cite{Lindstrom:2005zr}. Earlier relevant discussions may be found in \cite{Sevrin:1996jq}-\cite{Bogaerts:1999jc}.

The $d=2, \N=(2,2)$ algebra of covariant derivatives is
\beqs
&\{\mathbb{D}_{\pm},\bar{\mathbb{D}}_{\pm}\}=\pm i\pa_{\pp}~,
&\{\mathbb{D}_{\pm},\mathbb{D}_{\pm}\}=0\cr
&\{\mathbb{D}_{\pm},\mathbb{D}_{\mp}\}=0~,~~~~~~
&\{\mathbb{D}_{\pm},\bar\mathbb{D}_{\mp}\}=0
\eeqs
A chiral superfield $\phi$ satisfies the same constraints as in $d=4$:
\be
\bar{\mathbb{D}}_{\pm}\phi=\mathbb{D}_{\pm}\bar\phi=0~,
\ee
but in $d=2$ we may also introduce twisted chiral fields $\chi$ that satisfy
\beqs
&&\bar{\mathbb{D}}_{+}\chi=\mathbb{D}_{+}\bar\chi=0~,\cr
&&{\mathbb{D}}_{-}\chi=\bar{\mathbb{D}_{-}}\bar\chi=0~.
\eeqs
The sigma model action
\be
S= \int d^2\theta d^2\bar\theta K(\phi,\bar\phi,\chi,\bar\chi)~,
\ee
then precisely yields the GKG on $ker[J^{(+)},J^{(-)}]$, as may be seen by reducing the action to a $\N=(1,1)$-formulation. Denoting the $(1,1)$ covariant deivatives by $D_\pm$ and the generators of the second supersymmetry $Q_\pm$ we have
\beqs\label{reduce}
&&D_{\pm}:= \frac 1 {\sqrt 2}(\mathbb{D}_{\pm}+\bar{\mathbb{D}}_{\pm})\cr
&&Q_{\pm}:= i\frac 1 {\sqrt 2}(\mathbb{D}_{\pm}-\bar{\mathbb{D}}_{\pm})~.
\eeqs
Note that this formulation shows that both the metric and the $H$-field have $K$ as a potential
\beqs\label{kapot}\nonumber
&&G_{\phi\bar\phi}=K_{\phi\bar\phi}~, \quad G_{\chi\bar\chi}=-K_{\chi\bar\chi}\\[1mm]
&&H=\bar\pa_{\chi}\pa\_{\phi}\bar\pa_{\phi} K+\pa_{\chi}\pa_{\phi}\bar\pa_{\phi} K
+\bar\pa_{\phi}\pa_{\phi}\bar\pa_{\chi} K-\pa_{\phi}\pa_{\phi}\bar\pa_{\chi} K~,
\eeqs
where derivatives on $K$ are understood in the first line, and the second line is a three-form written in terms of the holomorphic differentials:
\be
d=\pa_{\phi}+\bar\pa_{\phi}+\pa_{\chi}+\bar\pa_{\chi}~.
\ee
See \cite{Ivanov:1994ec} for a more detailed discussion of the above coordinatization.

Furthermore, as discussed for GKG in the previous section, the commuting complex structures imply the existence of a local product structure
\be
\mathfrak{\mathfrak{P}}=-J^{(+)}J^{(-)},\quad \na {\mathfrak{P}}=0,\quad {\mathfrak{P}}^2=1~.
\ee
To discriminate it from the general GK case we call this a ``Bihermitean Local Product'' (BiLP) geometry.

The general case with $(ker[J^{(+)},J^{(-)}])^\bot\ne \emptyset $ was long a challenge\footnote{In a number of publications co-authored by me, $(ker[J^{(+)},J^{(-)}])^\bot$ was was incorrectly denoted $coker[J^{(+)},J^{(-)}]$. I apologize for participating in this misuse.}. The key issue here is what addidtional  
$\N=(2,2)$ superfields (if any) would suffice to describe the geometry. The available fields are complex linear $\Sigma_\phi$, twisted complex linear $\Sigma_\chi$ and semichiral superfields. Of these $\Sigma_\phi$ are dual to chirals and $\Sigma_\chi$ to twisted chirals (See appendix A).
The candidate superfields are thus  left and right semi-(anti)chirals $\mathbb{X}_{L.R}$ which obey
\beqs
&&\bar{\mathbb{D}}_{+}\mathbb{X}_{L}={\mathbb{D}}_{+}\bar{\mathbb{X}}_{L}=0~,\cr
&&\bar{\mathbb{D}}_{-}\mathbb{X}_{R}={\mathbb{D}}_{-}\bar{\mathbb{X}}_{L}=0~.
\eeqs
A $\N=(2,2)$ model written in terms of these fields reads
\be
S= \int d^2\theta d^2\bar\theta K(\mathbb{X}_{L.R},\bar{\mathbb{X}}_{L.R})~,
\ee
and an equal number of left and right fields are needed to yield a sensible sigma model. 
When reduced to $\N=(1,1)$ superspace, this action gives a
more general model than what we have considered so far. Using (\ref{reduce}) we have the following  $\N=(1,1)$ superfield content;
\beqs\label{comps}
&&X_{L}\equiv \mathbb{X}_{L}|, \quad \Psi_{L-}\equiv Q_{-}\mathbb{X}_{L}|,~\cr
&&X_{R}\equiv \mathbb{X}_{R}|, \quad \Psi_{R+}\equiv Q_{+}\mathbb{X}_{R}|~,
\eeqs
where the vertical bar now denotes setting {\em half} the fermi-coordinates to zero. Clearly, $X_{L,R}$ are scalar superfields and hence suitable for
the  $\N=(1,1)$ sigma model, but $\Psi_{L,R\pm}$ are spinorial fields. They enter the reduced action as auxiliary fields and are the auxiliary
$\N=(1,1)$ superfields needed for closure of the $\N=(2,2)$ algebra when $[J^{(+)},J^{(-)}]\ne 0$ (see \cite{Buscher:1987uw}). The structure of such an 
a $\N=(1,1)$ action is schematically \cite{Lindstrom:2004eh}
\be
S= \int d^2\theta d^2\bar\theta(D_{+}XE^{-1}D_{-}X+D_{(+}X\Psi_{-)})~,
\ee
where $E=G+B$ as before.

In \cite{Lindstrom:2005zr} we show that a sigma model fully describing GKG, i.e.,
$ker[J^{(+)},J^{(-)}]\oplus (ker[J^{(+)},J^{(-)}])^\bot$, is\footnote{Away from irregular points, i.e., points where the Poisson structures (\ref{p1}), (\ref{p3}) change rank.}
\be\label{gkact}
S= \int d^2\theta^2d^2\bar\theta K(\mathbb{X}_L,\bar\mathbb {X}_L, \mathbb{X}_R,\bar\mathbb{X}_R,\phi,\bar\phi,\chi,\bar\chi)~,
\ee
where $K$ acts as a generalized K\"ahler potential in terms of derivatives of which all geometric quantities can be expressed (locally). This
$K$ has the additional interpretation as a generating function for symplectomorphisms between certain sets of coordinates on
$(ker[J^{(+)},J^{(-)}])^\bot$, the canonical coordinates for $J^{(+)}$ and $J^{(-)}$, respectively. The proof of these statements relies heavily on Poisson geometry  \cite{Lindstrom:2005zr} and is summarized in what follows.

First, the fact that $(\phi,\chi)$ and their hermitean conjugates are enough to describe $ker[J^{(+)},J^{(-)}]$ may be reformulated using the 
Poisson-structures \cite{Lyakhovich:2002kc}
\be\label{p1}
\pi_\pm \equiv(J^{(+)}\pm J^{(-)})G^{-1}ª~.
\ee
In a neighbourhood of a regular point, coordinates may be choosen such that
\beqs\label{p2}
&& \pi^{A\mu}=0, \Rightarrow J^{(+)A}_{\mu}=J^{(-)A}_{\mu}\cr
&&\cr
&& \pi^{A'\mu}=0, \Rightarrow J^{(+)A'}_{\mu}=-J^{(-)A'}_{\mu}~.\cr
&&
\eeqs
It can be shown that $A\ne A'$ and that we have coordinates labeled $(a,a',A.A')$
adapted to 
\be
ker (J^{(+)} - J^{(-)}) \oplus ker (J^{(+)} + J^{(-)}) \oplus  (ker[J^{(+)},J^{(-)}])^\bot~,
\ee
where
\be\label{picoords}
J^{\pm }= \left( \begin{array}{cccc}
* & * &  * &  *\\
**&*& *& *\\
0 & 0 & I_c & 0\\
0 & 0 & 0& \pm I_t
\end{array} \right)~.
\ee
Here $I_{c}$ and $I_{t}$ have the canonical form
\be\label{cancomp}
I= \left( \begin{array}{cc}
i & 0\\
0&-i
\end{array} \right)
\ee
We thus have nice coordinates for $ker (J^{(+)} - J^{(-)}) \oplus ker (J^{(+)} + J^{(-)}) =ker[J^{(+)},J^{(-)}]$, but
$(ker[J^{(+)},J^{(-)}])^\bot$ remains to be described. Here a third Poisson structure turns out to be useful;
\be\label{p3}
\sigma\equiv  [J^{(+)},J^{(-)}]G^{-1}=\pm  (J^{(+)} \mp J^{(-)})\pi_{\pm}~.
\ee
Now $ker~\sigma=ker~\pi_{+}\oplus ker~\pi_{-}$ so we focus on  $(ker~\sigma)^\bot$. The symplectic leaf for $\sigma$ is $(ker[J^{(+)},J^{(-)}])^\bot$ 
and the third Poisson structure also has the following 
useful properties \cite{Hitchin}:
\beqs\label{sigmarln}
&&J^{\pm }\sigma J^{\pm t}=-\sigma~, \cr
&&\sigma = \sigma^{(2,0)} + \bar{\sigma}^{(0,2)}~,\cr
&&\bar\pa \sigma^{(2,0)}=0~,
\eeqs
where the holomorphic types are with respect to {\em both} complex structures.
To investigate the consequences of (\ref{sigmarln}) it is advantageous to first consider the case when
$ker[J^{(+)},J^{(-)}]=\emptyset$. It then follows that $\sigma$ is invertible and its inverse $\Omega$ is a symplectic form;
\be
\Omega = \sigma ^{-1}, \quad d\Omega=0~.
\ee
We may chose coordiantes adapted to $J^{(+)}$
\be
J^{(+)} = \left( \begin{array}{cc}
I_s & 0\\
0 & I_s
\end{array} \right)~.
\ee
In those coordinates we have from  (\ref{sigmarln}) that
\beqs\nonumber
&&\Omega =\Omega_{(+)}^{(2,0)}+\bar\Omega_{(+)}^{(0,2)}~,\\[1mm]
&&\pa\Omega_{(+)}^{(2,0)}=0=\bar\pa\bar\Omega_{(+)}^{(0,2)}~,
\eeqs
which identifies $\Omega_{(+)}^{(2,0)}$ as a holomorphic symplectic structure. 
The coordinates may then be further specified to be Darboux coordinates for this symplectic structure
\be
\Omega=dq^{a}\wedge dp^{a} + c.c.
\ee
The same derivation with $J^{(+)}$ replaced by $J^{(-)}$ gives a second set of Darboux coordinates which are  
canonical coordinates for $J^{(-)}$ and where
\be
\Omega=dQ^{a'}\wedge dP^{a'} + c.c.
\ee
Clearly the two sets of canonical coordinates are related by a symplectomorphism. Let $K(q,P)$ denote a generating function for 
this symplectomorphism. Expressing all our quantities 
in the mixed coordinates $(q,P)$, we discover that the expressions for $J^{(\pm)} , ~\Omega, G=\Omega [J^{(+)}, J^{(-)}],..$ are precicesly what 
we\footnote{See also \cite{Buscher:1987uw}\cite{Bogaerts:1999jc} for partial results.} derived from the sigma model action (\ref{gkact}) provided that we identify the coordinates $(q,P)$ with 
$(\mathbb{X}_{L},\mathbb{X}_{R})$ (and the same for the hermitean conjugates.)

In the general case when $[J^{(+)}, J^{(+)}]\ne \emptyset$, we again get agreement, provided that the coordinates indexed
$A$ and $A'$ in (\ref{picoords}) are identified with the chiral and twisted chiral fields $(\phi,\chi)$. We thus have a one to 
one correspondence between the descripton covered by the sigma model and all of $[J^{(+)}, J^{(+)}]$, i.e., for all possible cases.

\section{Linearization of Generalized K\"ahler Geometry}
\setcounter{equation}{0}

The generalized \Ka potential 
$K(\mathbb{X}_L,\bar\mathbb {X}_L, \mathbb{X}_R,\bar\mathbb{X}_R,\phi,\bar\phi,\chi,\bar\chi)$
yields all geometric quantities, but as non-linear expressions\footnote{Dualizing a BiLP to (twisted) complex linear fields yields a model with similar nonlinearities.} in derivatives of $K$ (unlike the case (\ref{kapot})).
In \cite{Lindstrom:2007qf} we show that these non-linearities can be viewed as arising from a quotient of a higher dimensional model with certain null
Kac-Moody symmetries. To illustrate the idea, we first consider an example.
\subsection{A bosonic example.}
\label{boson}

Consider a Lagrangian of the form
\be\label{l1}
L_1:=A^\mu A_\mu~.
\ee
Following St\"uckelberg, we may think of this as a gauge fixed version of the gauge-invariant  Lagrangian
\be\label{l2}
L_2:=D^\mu\vf D_\mu\vf~,
\ee
with $D_\mu\vf:=\pa_{\mu}\vf +A_\mu$ and gauge invariance
\beqs\label{gi}
&\delta \vf =\epsilon\cr
&\delta A_\mu=-\pa_{\mu}\epsilon~.
\eeqs
Finally, the Lagrangian $L_2$ can in turn be thought of as arising through  gauging of the global translational symmetry $\delta \vf =\epsilon$ in a third Lagrangian
\be\label {l3}
L_3:=\pa{}^\mu\vf\pa_{\mu}\vf~.
\ee
A slightly more elaborate example is provided by the following sigma model Lagrangian;
\be\label{l4}
\tilde L_1:=G_{ab}(\phi)\pa{}^\mu\phi^a\pa{\mu}\phi^b+2G_a(\phi)\pa{}^\mu\phi^aA_\mu+G(\phi)A^\mu A_\mu~,
\ee
where $A_\mu$ is an auxiliary field.  Following the line of reasoning above, this Lagrangian may be thought of as a gauge fixed version of
\be
\tilde L_2:=G_{ab}(\phi)\pa{}^\mu\phi^a\pa{\mu}\phi^b+2G_{a0}(\phi)\pa{}^\mu\phi^aD_\mu\vf
+G_{00}(\phi)D^\mu\vf D_\mu\vf~,
\ee
where $D_\mu$ is as defined above, $G_{a0}\equiv G_a,~ G_{00}\equiv G$ and the St\"uckelberg field $\vf\equiv \phi^0$. In turn $\tilde L_2$ is the gauged version (in adapted coordinates) of the Lagrangian
\be\label{l6}
\tilde L_3:=G_{ij}(\phi)\pa{}^\mu\phi^i\pa_{\mu}\phi^j~,\quad i=0,a~,
\ee
whose global symmetry is given by the isometry
\be\label{is}
\pa_{0}G_{ij}=0~.
\ee
We see that eliminating the auxiliary field in $\tilde L_1$ is tantamount to extremizing $\tilde L_2$ with respect to the gauge field, i.e., to constructing a quotient of the Lagrangian $\tilde L_3$ with respect to its isometry. The only remaining question seems to be if varying the gauge-fixed $\tilde L_1$ is the same as varying $\tilde L_2$ (modulo gauge-fixing). The resulting $A_\mu$'s differ by a gauge-transformation $\pa_{\mu}\vf$. Explicitly:
\beqs\label{gt2}
&&\delta \tilde L_1=0 \Rightarrow -A_\mu =\frac  {G_a\pa_{\mu}\phi^a} G~,\cr
&&~\cr
&&\delta \tilde L_2=0 \Rightarrow -(\pa_{\mu}\vf+A_\mu) =\frac  {G_{a0}\pa_{\mu}\phi^a} {G_{00}}~.
\eeqs

\subsection{The Generalized K\"ahler Potential}

We  apply the procedure described above to a semichiral sigma model.  Most of the rest of this section is taken directly from \cite{Lindstrom:2007qf} and
\cite{Lindstrom:2007xv} where more details may be found.

Consider the generalized K\"ahler potential
\be\label{kap}
K(\bbX{L,R},\bbXB{\bar L,\bar R})~,
\ee
where $\bbX{L,R}$ are left and right semi-chiral $N=(2,2)$ superfields:
\be\label{sem}
\bbDB{+}\bbX{L}=0=\bbDB{-}\bbX{R}~.
\ee
We descend to $N=(1,1)$ as in (\ref{comps}) by defining  components
\beqs\nonumber
X_L = \mathbb{X}_L|,\;\;\;
\Psi_{L-} = Q_-\mathbb{X}_{L}|,\\
X_R = \mathbb{X}_R|,\;\;\;
\Psi_{R+} = Q_+\mathbb{X}_R|,
\eeqs
which satisfy
\beqs\label{qs}
Q_+\Psi = J D_+\Psi_{L-},\;\;\;
Q_-\Psi_{L-} = -i\partial_= X_L,\\
Q_-\Psi_{R+}  = JD_-\Psi_{R+},\;\;\;
Q_+\Psi_{R+} = -i\partial_= X_R.
\eeqs
The $N=(1,1)$ form of the Lagrangian is
\beqs\label{tjo}
\left(\begin{array}{cccc}
D_+X_L &
\Psi_{L+} & D_+X_R &
\Psi_{R+}\end{array}\right) E
\left(\begin{array}{c}
D_-X_L\\
\Psi_{L-}\\
D_-X_R\\
\Psi_{R-}\end{array}\right)
\eeqs
where
\beqs\label{semiE}
E =
\left(\begin{array}{cccc}
0 & K_{LL} + IK_{LL}I &
IK_{LR}J & 0\\
0 & 0 & 0 & 0\\
0 & K_{RL} & 0 & 0\\
K_{RL} & IK_{RL}I & K_{RR}+IK_{RR}I & 0
\end{array}\right)~.
\eeqs
Here we use a short hand notation where, e.g., $K_{LR}$ denotes the matrix of second derivatives of the potential
(\ref{kap}) with respect to both bared and un-bared  left and right fields. Also the canonical complex  structures $I$ is defined in (\ref{cancomp}).
Notice that neither $\Psi_{L+}$ nor $\Psi_{R-}$ occur in the action.

\subsection{ALP and Kac-Moody quotient}

The procedure $\tilde L_1\to\tilde L_2\to\tilde L_3$ outlined in Sec.\ref{boson} applied to the present case entails the replacements $(\Psi_{L-},\Psi_{R+})\to (\hcd{-}\vf_L,\hcd{+}\vf_R)\to(D_-\vf_L,D_+\vf_R)$ with $\tilde L_1$ given by  (\ref{tjo}), and where
\beqs\label{nab}
\hcd{\pm}\vf_{R/L}:=D_\pm\vf_{R/L}+\Psi_{R/L \pm}~.
\eeqs\label{ginv}
The gauge invariance of $\tilde L_2$ is 
\beqs
\delta \vf_{R/L}= \varepsilon_{R/L}~, ~~\delta\Psi_{R/L \pm}=-D_\pm \varepsilon_{R/L}~,
\eeqs
which gauges the following ``global''  invariance of $\tilde L_3$:
\beqs\label{glinv}
\delta \vf_{R/L}= \varepsilon_{R/L}~, ~~D_\pm \varepsilon_{R/L}=0~.
\eeqs
Since the matrix $E$ in (\ref{semiE}) is independent of $\vf_{R/L}$, invariance under (\ref{ginv}) and (\ref{glinv})
is immediate. Using the metric $G:=\frac 1 2 (E+E^t)$ one also verifies that the Killing-vectors
\beqs\label{km}
k_R:=\left(\begin{array}{c}
0\cr
0\cr
0\cr
\varepsilon_R\end{array}\right)~,\qquad k_L:= \left(\begin{array}{c}
0\cr
\varepsilon_L\cr
0\cr
0\end{array}\right)~,
\eeqs
are null-vectors. Furthermore, the constraints in (\ref{glinv}) imply
\beqs\label{KM}
&D_+k_R=0&,\cr
&D_-k_L=0&,\cr
&\partial_{\+}k_R=0&,\cr
&\partial_{=}k_L=0&~,
\eeqs
or covariantly  \cite{Rocek:1991ps}
\be
\nabla^{(+)}k_{LA}=0=\nabla^{(-)}k_{RA}~,
\ee
with $\nabla^{(\pm)}$ defined in  (\ref{nablator}).
These relations identify the global symmetries as null Kac-Moody isometries.\\

After applying the procedure outlined above, we obtain a Lagrangian $\tilde L_3$. It is then useful to introduce a definition from \cite{Lindstrom:2007qf}:\\

\noindent
{\em The space corresponding to $\tilde L_3$  is the $N=(1,1)$ form of the Auxiliary Local Product space (ALP) for the 
$N=(2,2)$ Lagrangian  $\tilde L_1$ in (\ref{tjo}).}\\

\noindent
 In other words, the $ALP$ is given by the action
\beqs\label{tjosan}
\left(\begin{array}{cccc}
D_+X_L &
\bullet& D_+X_R &
D_{+}\vf_R\end{array}\right) E
\left(\begin{array}{c}
D_-X_L\\
D_-\vf_{L}\\
D_-X_R\\
\bullet \end{array}\right)~,
\eeqs
where $E$ is the matrix given in (\ref{semiE}), the bullets denote the decoupled $\Psi_{L+}, \Psi_{R-}$, and the Lagrangian is invariant under the global Kac-Moody isometry (\ref{km}). 

\subsubsection{Kac-Moody quotient in $(1,1)$}
The Lagrangian (\ref{tjosan}) is an equivalent starting point for deriving the GK geometry for the target space of  (\ref{tjo}): To recapitulate from Sec.\ref{boson}, this proceeds by gauging the isometry to obtain the $\tilde L_2$ Lagrangian
\beqs\label{hejsan}
\left(\begin{array}{cccc}
D_+X_L &
\bullet& D_+X_R &
D_{+}\vf_R+\Psi_{R+}\end{array}\right) E
\left(\begin{array}{c}
D_-X_L\\
D_-\vf_{L}+\Psi_{L-}\\
D_-X_R\\
\bullet \end{array}\right)~.
\eeqs\label{psieqns}
Elimination of the gauge fields (cf (\ref{gt2}));
\beqs
&\delta\Psi_{L-} \Rightarrow D_-\vf_L+\Psi_{L-}=-JK^{-1}_{LR}J(K_{RR}+JK_{RR}J)D_-\bbX{R}-JK^{-1}_{LR}JK_{RL}D_-\bbX{L}\cr
&~\cr
&\delta\Psi_{R+} \Rightarrow D_-\vf_R+\Psi_{R+}= -JK^{-1}_{RL}J(K_{LL}+JK_{LL}J)D_-\bbX{L}-JK^{-1}_{RL}JK_{LR}D_-\bbX{R}~,\cr
& ~\cr
&~
\eeqs
 yields the quotient metric and $B$-field from $\mathbf{ E}$:
\beqs\label{moon}
&\mathbf{E}=
\left(\begin{array}{cc}
 C_{LL}K_{LR}^{-1}JK_{RL}& JK_{LR}J + C_{LL}K_{LR}^{-1}C_{RR} \\
 -K_{RL}J K_{LR}^{-1} J K_{RL} & -K_{RL}J K_{LR}^{-1} C_{RR}\\
\end{array}\right)~,\cr
&~
\eeqs
where, supressing indices on the two by two complex matrices,
\beqs\label{c}
C:=[J,K]~.
\eeqs
The corresponding Lagrangian is
\beqs\label{tjolahopp}
\left(\begin{array}{cccc}
D_+X_L &
 D_+X_R 
\end{array}\right) \mathbf{E}
\left(\begin{array}{c}
D_-X_L\\
D_-X_R\\
\end{array}\right)
\eeqs

\subsubsection{Kac-Moody quotient in $(2,2)$}
As an alternative, we may perform the Kac-Moody quotient in $(2,2)$ superspace. Very briefly, this goes as follows:

In the generalized potential we replace the semi-chiral fields by sums of chiral and twisted chiral fields according to
\beqs\label{alp}\nonumber
&&K(\mathbb{X}_L,\bar\mathbb {X}_L, \mathbb{X}_R,\bar\mathbb{X}_R,\phi,\bar\phi,\chi,\bar\chi)~,\\[1mm]
\longrightarrow &&K(\phi_L+\chi_{L},\bar\phi_L+\bar\chi_{L}, \phi_R+\bar\chi_{R},
\bar\phi_R+\chi_{R},....)~.
\eeqs
This doubles the degrees of freedom in the semi sector but the corresponding action has a Kac-Moody symmetry
\beqs
&&\delta\phi_{L,R}=\lambda_{L,R}~,\cr
&&\delta\chi_{L}=-\lambda_{L}~,\cr
&&\delta\chi_{R}=-\bar\lambda_{R}~,
\eeqs
where the parameters satisfy
\beqs\nonumber
&&\bar{\mathbb{D}}_{\pm}\lambda_{L}=\mathbb{D}_{-}\lambda_{L}=0~,\quad \pa_{=}\lambda_{L}=0~,\\[1mm]
&&\bar{\mathbb{D}}_{\pm}\lambda_{R}=\mathbb{D}_{+}\lambda_{R}=0~,\quad \pa_{\+}\lambda_{R}=0~,
\eeqs
To keep the same degrees of freedom as in the original model, we gauge the Kac-Moody symmetry which reintroduces semi-chiral fields:
\beqs
&&K(\phi_L+\chi_{L},\bar\phi_L+\bar\chi_{L}, \phi_R+\bar\chi_{R},
\bar\phi_R+\chi_{R},....)~.\cr
\longrightarrow &&K(\phi_L+\chi_{L}+\mathbb{X}_L,\bar\phi_L+\bar\chi_{L}+\bar\mathbb {X}_L, 
\phi_R+\bar\chi_{R}+\mathbb{X}_R,\bar\phi_R+\chi_{R}+\bar\mathbb{X}_R,...)~.
\eeqs
The local complex Kac-Moody symmetry is now
\beqs
&&\delta\phi_{L,R}=\Lambda_{L,R}~,\cr
&&\delta\chi_{L}=-\tilde\Lambda_{L}~,\cr
&&\delta\chi_{R}=-\bar{\tilde{\Lambda_{R}}}~,\cr
&&\delta\mathbb{X}_{L,R}=-\Lambda_{L,R}+\tilde\Lambda_{L,R}~.
\eeqs
The equivalence to the generalized potential is seen by going to a gauge where the ``$\phi+\chi$'' terms are zero.

For  comparizon, we descend to $\N=(1,1)$ via the identification
\beqs
&&X_{L}=(\phi_{L}+\chi_{L})|~,\quad \psi_{L-}=Q_{-}(\phi_{L}+\chi_{L})|=ID_{-}(\phi_{L}-\chi_{L})\equiv ID_{-}\varphi_{L}\cr
&&X_{R}=(\phi_{R}+\bar\chi_{R})|~,\quad \psi_{R+}|\equiv ID_{+}\varphi_{R}~.
\eeqs
This gives the $\N=(1,1)$ action in terms of the complex scalar fields $X_{L,R}$ and $\varphi_{L,R}$, with the Kac-Moody generated by the null Killing 
vectors (\ref{km}).
These corresponding isometies may be quotiented to give precisely the nonlinear expressions in terms of derivatives of $K$ that we found in (\ref{moon}) .
They arise from
\be
\mathbf{ E}_{\mu\nu}=E_{\mu\nu}-k_{L\mu }h^{-1}k_{R\nu}, \quad h\equiv k_{R}Gk_{L}~,
\ee
where $E_{\mu\nu}$ are the $X_{L,R}$  components in (\ref{semiE}). Note that the existence of a left and a right isometry generalizes the construction in
(\ref{actare}) slightly, to allow for a $B$-field.

\section{Projective Superspace}
\setcounter{equation}{0}

Typically, the ${\cal{N}}=(2,2)$ formulation of the ${\cal{N}}=(4,4)$ models require explicit transformations on the ${\cal{N}}=(2,2)$ superfields that close to the supersymmetry algebra on-shell. This non-manifest formulation makes the construction of new models difficult. Below follows a brief description of a superspace where all supersymmetries are manifest. This  projective superspace\footnote{The name refers to the projective coordinates on $\mathbb{CP}^1=:\mathbb{P}^1$ being used. It is really a misnomer in that it is unrelated to the usual definition of projective spaces.} \cite{Karlhede:1984vr}-\cite{Gonzalez-Rey:1997db} has been developed 
independent of an in parallel to harmonic superspace 
\cite{Galperin:2001uw}. The relation between the two approaches was first discussed in \cite{Kuzenko:1998xm} and more recently in \cite{Jain:2009aj}.
A key reference for this section is \cite{Lindstrom:2008gs} and the review \cite{Kuzenko:2010bd}.

A hyperk\"ahler space ${\cal T}$ supports three globally defined integrable complex structures $I,J,K$ obeying the quaternion algebra: $IJ=-JI=K$, plus cyclic permutations. Any linear combination of these,  $aI+bJ+cK$ is again a complex structure on ${\cal T}$ if $a^2+b^2+c^2=1$, i.e., if $\{a,b,c\}$ lies on a two-sphere $S^2\backsimeq \mathbb{P}^1$.
The Twistor space $\Z$ of  a hyperk\"ahler space ${\cal T}$ is the product of ${\cal T}$ with this two-sphere $\Z= {\cal T}\times\mathbb{P}^1$. 
The two-sphere  thus parametrizes the complex structures and we choose projective coordinates $\zeta$ to describe it (in a patch including the north pole). It is an interesting and remarkable fact that the very same
$S^2$ arises in an extension of superspace to accomodate manifes ${\cal{N}}=(4,4)$ models.

Although projective superspace can be defined for different bosonic dimensions, we shall remain in two. Here the algebra of ${\cal{N}}=(4,4)$ superspace derivatives is
\beqs
&\{\mathbb{D}_{a\pm},\bar{\mathbb{D}}^b_{\pm}\}=\pm i\delta^b_a\pa_{\pp}~,
&\{\mathbb{D}_{a\pm},\mathbb{D}_{b\pm}\}=0\cr
&\{\mathbb{D}_{a\pm},\mathbb{D}_{b\mp}\}=0~,~~~~~~
&\{\mathbb{D}_{a\pm},\bar\mathbb{D}^b_{\mp}\}=0
\eeqs
We may parameterize a \cpl
of maximal graded abelian subalgebras
as (suppressing the spinor indices)
\be
\label{projder}
\nabla(\z)={\mathbb{D}}_{2}+\z {\mathbb{D}}_{1}~,~~
\bar{\nabla} (\z)
=\bar{\mathbb{D}}^1-\z\bar{\mathbb{D}}^2~,
\ee
where $\z$ is the coordinate introduced above, and the bar on $\nabla$ denotes conjugation with respect to a real structure $\mathfrak{R}$ defined as complex conjugation composed with the antipodal map on
$\mathbb{P}^1\backsimeq S^2$. 
The two new covariant derivatives in (\ref{projder}) anti-commute
\be
\{\nabla ,\bar \nabla\}=0~.
\ee
They may be used to introduce constraints on superfields similarily to how the ${\cal{N}}=(2,2)$ derivatives are used to impose chirality constraints in  Sec.\ref{offshell}. Superfields now live in an  extended superspace with coordinates $\xi,\z, \theta$. The superfields $\Y$ we shall be interested in satisfy the projective chirality constraint
\be
\label{projchir}
\nabla \Y=\bar \nabla\Y=0~,
\ee
and are taken to have the following $\z$-expansion:
\be
\label{expansion1}
\Y=\sum_i\Y_i\z^i~.
\ee
When the the index $i\in[0,\infty)$ the field $\Y$ is analytic around the north pole of the $\mathbb{P}^1$ and consequently called an arctic multiplet.
For tropical and antarctic multiplets see \cite{Gonzalez-Rey:1997qh}.
We use the real structure  acting on superfields, $\mathfrak{R}(\Y)\equiv \bar\Y$, to impose reality conditions on the superfields. An $\OO (2n)$ multiplet is thus defined via
\be
\label{otwonmult}
\Y\equiv \eta_{(2n)}=(-)^n\z^{2n}\bar\Y~.
\ee

The expansion (\ref{expansion1}) is useful in displaying the ${\cal{N}}=(2,2)$ content of the multiplets.
Using the relation (\ref{projder}) to the ${\cal{N}}=(2,2)$ derivatives in (\ref{projchir}) we read off the following expansion for an $\OO (4)$ multipet (\ref{otwonmult}):
\be
\eta_{(4)}=\phi +\z \Sigma +\z^2X-\z^3\overline{\Sigma} + \z^4\bar\phi~,
\ee
with the component ${\cal{N}}=(2,2)$ fields being chiral $\phi$, unconstrained $X$ and complex linear $\Sigma$.
A complex linear field satisfies
\be
\bar{\mathbb{D}}^2\Sigma =0~,
\ee
and is dual to a chiral superfield (see the appendix).
A general arctic projective chiral $\Y$ has the expansion
\be\label{arctic}
\Y=\phi +\z \Sigma +\sum_{i=2}^\infty X_i\z^i~,
\ee
with all $X_i$'s unconstrained.

\subsection{The Generalized Legendre Transform}

In this section we review one particular construction of hyperk\"ahler metrics using projective superspace  introduced in \cite{Lindstrom:1987ks}.

An ${\cal{N}}=(4,4)$ invariant action for the field in (\ref{arctic}) may be written as
\be
\label{act1}
S=\int\mathbb{D}^2\bar{\mathbb{D}}^2F~,
\ee
with 
\be
\label{action2}
F\equiv \oint_C\frac{d\z }{2\pi i \z }f(\Y,\bar\Y;\z)~,
\ee
for some suitably defined contour $C$.  Eliminating the auxiliary fields $X_i$ by their equations of motion will yield an ${\cal{N}}=(2,2)$ model defined on the tangent bundle 
$T(\cal{T})$ parametrized by $(\phi,\Sigma)$. Dualizing the complex linear fields $\Sigma$ to chiral fields $\tilde \phi$ the final result is a supersymmetric ${\cal{N}}=(2,2)$ sigma model in terms of $(\phi, \tilde\phi)$ which is guaranteed by construction to have ${\cal{N}}=(4,4)$ supersymmetry, and thus to define a hyperk\"ahler metric. In equations, these steps are:\\
Solve the equations of motion for the auxiliary fields:
\be\label{dF}
\frac{\pa F}{\pa\Y_i}=
\oint_C\frac{d\z}{2\pi i\z}\,\z^i
\left(\frac\pa{\pa\Y}f(\Y,\bY;\z)\!\right)=0~~,~~~i\ge 2~.
\ee
Solving these equations puts us on ${\cal{N}}=2$-shell, which means that only the ${\cal{N}}=(2,2)$ component symmetry remains off-shell. (In fact, insisting on keeping the ${\cal{N}}=(4,4)$ constraints
(\ref{projchir}) will put us totally on-shell.)
In ${\cal{N}}=(2,2)$ superspace the resulting model, after eliminating $X_i$, is given by a Lagrangian
$K(\phi,\bar\phi,\Sigma,\bar \Sigma)$.  This is finally dualized to $\tilde{K}(\phi,\bar \phi,\tilde \phi,\bar {\tilde \phi})$ via a Legendre transform
\beqs
&&\tilde{K}(\phi,\bar \phi,\tilde \phi,\bar{\tilde \phi})=K(\phi,\bar \phi,\Sigma,\bar \Sigma)-\tilde \phi \Sigma-\bar{\tilde \phi} \bar \Sigma\cr
&&\cr
&&\tilde \phi=\frac{\pa K}{\pa \Sigma}~,~~~\bar{\tilde \phi}=\frac{\pa K}{\pa \bar \Sigma}~.
\eeqs          

\subsection{Hyperk\"ahler metrics on Hermitean symmetric spaces}

 This section contains an introduction to  \cite{Arai:2006gg} where the generalized Legendre transform described in the previous section is used to find metrics on the Hermitean symmetric  spaces listed in the following table:

\bigskip
  \begin{center}
  \begin{tabular}[htb]{|l|l|}
  \hline
  Compact&Non-Compact\cr
  \hline
  $U(n+m)/U(n)\times U(m)$&$U(n,m)/U(n)\times U(m)$\cr
  $SO(2n)/U(n);~Sp(n)/U(n)$&$SO^*(2n)/U(n);~Sp(n,\mathbb{R})/U(n)$\cr
  $SO(n+2)/SO(n)\times SO(2)$& $SO_0(n+2)/SO(n)\times SO(2)$\cr
  \hline
  \end{tabular}
  \end{center}
  \bigskip

The special features of these quotient spaces that allow us to find a hyperk\"ahler metric on their co-tangent bundle is the existence of holomorphic isometries and that we are able to find convenient coset representatives.

A simple example of how the coset representative enters in understanding a quotient is given, e.g., in  \cite{vanNieuwenhuizen:1984ke}: In $\mathbb{R}^{n+1}$
the sphere $S^n$ forms a representation of $SO(n+1)$. The isotropy subgroup at the north pole $p_0$ of $S^n$ is $SO(n)$. Consider another point $p$ on $S^n$ an let $g_p\in SO(n+1)$ be an element that maps $p_0 \to p$. The complete set of elements of  $SO(n+1)$ which map $p_0 \to p$ is thus of the form $g_pSO(n)$, or in other words $S^n=SO(n+1)/SO(n)$. A coset representative is a choice of element in  
 $g_pSO(n)$, and that choice can make the transport of properties defined at the north pole to an arbitrary point more or less transparent.

An important step in the generalized Legendre transform is to solve the auxiliary field equation (\ref{dF}). As outlined in \cite{GK1} and further elaborated in \cite{K3}, 
for Hermitian symmetric spaces 
the auxiliary fields may be eliminated exactly. In the present case, we start from a solution at the origin $\phi=0$,
\be
\label{auxi}
\Y^{(0)}=\z\Sigma^{(0)}~. 
\ee
We then extend this solution to a solution $\Y^*$ at an arbitrary point using a coset representative.
We illustate the method in an example due to S. Kuzenko.
\eject

{\em Ex. (Kuzenko)}

\begin{quotation}
The K\"ahler potential for \cpl is given by
\be
K(\phi,\bar \phi)=ln(1+\phi\bar \phi)~,
\ee
and we denote the metric that follows from this by $g_{\phi , \bar \phi}$. Here 
$\phi$ 
is a holomorphic coordinate which we extend to an $\N=(2,2)$ chiral superfield.
To construct a  hyperk\"ahler metric we first replace $\phi\to \Y$, and then solve the auxiliary field equation as in (\ref{auxi}). Thinking of 
$\mathbb{C}\mathbb{P} ^n$ as the quotient $G_{1,n+1}({\mathbb C})= {\rm U} (n+1) / {\rm U}(n) \times {\rm U}(1)$, we use a carefully chosen coset representative $L(\phi, \bar \phi)$ to extend the solution from the origin to an arbitrary point. The result is
\be
\Y^*=\frac {\Y^{(0)}+\phi }{1-\Y^{(0)}\bar \phi }=\frac {\z\Sigma^{(0)}+\phi }{1-\z\Sigma^{(0)}\bar \phi }~.
\ee
To find the chiral multiplet $\Sigma$ that parametrizes the tangent bundle, we use the definition
\be
\Sigma \equiv \frac{d\Y^*}{d\z }|_{\z=0}=(1+\phi\bar \phi)\Sigma^{(0)}~,
\ee
yielding
\be
\Y^*=\frac{(1+\phi\bar \phi)\phi+\z\Sigma }{(1+\phi\bar \phi)-\z\Sigma \bar\phi }~.
\ee
The $\N=(2,2)$ superspace Lagrangian on the tangent bundle is then
\be
K(\Y^*,\bar \Y^*)=K(\phi,\bar\phi)+ln(1-g_{\phi\bar\phi}\Sigma\bar \Sigma)~.
\ee   
The final Legendre transform replacing the linear multiplet by a new chiral field, $\Sigma \to \tilde \phi$   produces the K\"ahler potential $K(\phi,\bar \phi, \tilde \phi, \bar {\tilde \phi})$ for the Eguchi-Hanson metric.    \end{quotation}
\bigskip

The \cpl example captures the essential idea in our construction. The reader is referred to the papers \cite{Arai:2006gg}-\cite{Kuzenko:2008ci} for more examples. 

\subsection{Other alternatives in Projective Superspace} 

Of the two methods for constructing hyperk\"ahler metrics introduced in 
\cite{Lindstrom:1983rt}, we have dwelt on the Legendre transform  generalized to projective superspace. The hyperk\"ahler reduction discussed in Sec.\ref{HK}, may also be lifted to projective superspace. Both these methods involve only chiral  
$\N=(2,2)$ superfields. When a nonzero $B$-field is present, the $\N=(2,2)$ sigma models involve chiral, twisted chiral and semichiral superfields, as discussed in Sec.\ref{sigma}. For a full description of (generalizations of) hyperk\"ahler metrics on such spaces, the doubly projective superspace \cite{Buscher:1987uw} is required. We now briefly touch on this construction.

In the doubly projective superspace, at each point in ordinary superspace we intrduce one \cpl for each chirality and denote the corresponding coordinates by 
$\z_L$ and $\z_R$. The condition (\ref{projder}) turns into
\beqs
\label{LRprojder}
&&\nabla_+(\z_L)={\mathbb{D}}_{2+}+\z_L {\mathbb{D}}_{1+}~,\cr
&&\cr
&&\nabla_-(\z_R)={\mathbb{D}}_{2+}+\z_R {\mathbb{D}}_{1-}~,
\eeqs
with the conjugated operators defined with respect to the real structure   
$\mathfrak{R}$ acting on both $\z_L$ and $\z_R$. A superfield has the expansion
\be
\label{expansion2}
\Y=\sum_{i,j}\Y_{i,j}\z_L^i\z_R^j~,
\ee
and is taken to be both left and right projectively chiral. We may also impose reality conditions using $\mathfrak{R}$, as well as particular conditions on the components, such as the ``cylindrical'' condition
\be
\Y_{i,j+k}=\Y_{i,j}~,
\ee
for some $k$.
Actions are formed in analogy to (\ref{act1}) and (\ref{action2}). The $\N=(2,2)$ components of such a model  include twisted chiral fields $\chi$, as well as semi-chiral
ones $\mathbb{X}_{L,R}$. In fact this is the context in which the semi-chiral $\N=(2,2)$ superfields were introduced \cite{Buscher:1987uw}.
 Hyperk\"ahler metrics derived in this superspace are discussed in \cite{Lindstrom:1994mw}.
An exciting project is to merge this picture with the  results in \cite{Lindstrom:2007qf}.
\bigskip

\noindent{\bf\large Acknowledgement}:
\bigskip

I am very happy to acknowledge all my collaborators on the papers that form the basis of this presentation. In particular I am grateful for the many years of continuous collaboration with Martin Ro\v cek, my intermittent collaborations with Chris Hull, as well as  the also long but more  recent  collaborations with Sergei Kuzenko, Rikard von Unge and Maxim Zabzine. The figure from \cite{Hitchin:1986ea} is reproduced with permission from Springer Verlag.
The work was supported by  VR grant 621-2009-4066.

\appendix
\appendixpage
\addappheadtotoc
\section{Chiral-Complex linear duality}
\setcounter{equation}{0}
In two dimensions, chiral superfields (\ref{chiral}) obey
\be
\bbDB{\pm}\phi=0~,
\ee
and twisted chiral superfields $\chi$ obey
\be
\bbDB{+}\chi=\bbD{-}\chi=0
\ee
and the complex conjugate relations. They are related via Legendre transformations to complex linear $\Sigma_\phi$ and twisted complex linear $\Sigma_\chi$ superfields obeying
\be
\bbDB{}^2\Sigma_\phi=0=\bbDB{+}\bbD{-}\Sigma_\chi~,
\ee
and the complex conjugate relations. 

A parent action which relates a BiLP generalized K\"ahler potential $K(\phi,\bar\phi,\chi,\bar\chi)$ to its dual 
$U(\Sigma_\phi,\overline{\Sigma}_\phi,\Sigma_\chi,\overline{\Sigma}_\chi)$ is 
\be\label{LTF}
K(X,\bar X,Y,\bar Y)-\Sigma_\phi X-\overline{\Sigma}_\phi\bar X-\Sigma_\chi Y-\overline{\Sigma}_\chi\bar Y~.
\ee
Variation of the (twisted) complex linear fields constrains $X,\bar X, Y, \bar Y$ to be $\phi,\bar\phi,\chi,\bar\chi$ and 
$K(\phi,\bar\phi,\chi,\bar\chi)$ is recovered.
On the other hand, the  $X,\bar X, Y, \bar Y$ field equations are
\be
K_X-\Sigma_\phi=0~, \quad K_{\bar X}-\overline{\Sigma}_\phi=0~,\quad K_Y-\Sigma_\chi=0~,\quad K_{\bar Y}-\overline{\Sigma}_\chi=0~.
\ee
Assuming that they can be solved for $X,\bar X, Y, \bar Y$ as functions of the (twisted) complex linear fields we find the Legendre transformed potential
$U(\Sigma_\phi,\overline{\Sigma}_\phi,\Sigma_\chi,\overline{\Sigma}_\chi)$ when the solutions are plugged back into (\ref{LTF}).

The above discussion is purely local. To consider global issues, one must take into account gluing of the potential between patches. In the BiLP case the allowed change between patches ${\cal{O}}_a$ and ${\cal{O}}_b$ is given by holomorphic coordinate transformations the (generalized) K\"ahler gauge transformations. 

Let us look at K\"ahler gauge transformations, restricting to the case with no twisted chiral fields for simplicity.
We thus have
\beqs\label{6}
\tilde K(\phi,\bar\phi)= K(\phi,\bar\phi)-F(\phi)-\bar F(\bar \phi)~,
\eeqs
which via a holomorphic coordinate transformation $\phi'=F(\phi)$ is equivalent to
\beqs\label{66}
{\cal K}(\phi',\bar\phi')= K(F^{-1}(\phi'), \bar F^{-1}(\bar\phi'))-\phi'-\bar \phi'~.
\eeqs

One may ask what this freedom corresponds to in the dual model where no ambiguity of the same type exists. 

The dual to $K$ is found from the Legendre transform with parent action
\beqs\label{7}
K(X,\bar X)-\Sigma_\phi X-\overline{\Sigma}_\phi\bar X~, 
\eeqs
and reads
\be\label{U}
U(\Sigma_\phi,\overline{\Sigma}_\phi)=K\left(X(\Sigma_\phi,\overline{\Sigma}_\phi),\bar X(\Sigma_\phi,\overline{\Sigma}_\phi)\right)-\Sigma_\phi X(\Sigma_\phi,\overline{\Sigma}_\phi)-\overline{\Sigma}_\phi\bar X(\Sigma_\phi,\overline{\Sigma}_\phi)~,
\ee
after solving
\be\label{first}
K_X-\Sigma_\phi=0~, \quad K_{\bar X}-\overline{\Sigma}_\phi=0~.
\ee
The parent action to $\tilde K$ is best considered after the coordinate transformation (\ref{66}):
\beqs\label{7}\nonumber
&&{\cal K}(X,\bar X)-\Sigma_\phi X-\overline{\Sigma}_\phi\bar X\\[1mm]\nonumber
&&=K\left(F^{-1}(X), \bar F^{-1}(\bar X)\right)-X-\bar X-\Sigma_{\phi'} X-\overline{\Sigma}_{\phi'} \bar X\\[1mm]
&&=K\left(F^{-1}(X), \bar F^{-1}(\bar X)\right)-\Sigma'_{\phi'} X-\overline{\Sigma}'_{\phi'} \bar X~,
\eeqs
where $\Sigma'_{\phi'}:=1+\Sigma_{\phi'}$. We find the corresponding dual potential
\beqs\nonumber\label{calU}
&&{\cal U}(\Sigma_{\phi'},\overline{\Sigma'}_{\phi'})\\[1mm]\nonumber
&&=K\left( F^{-1}(X(\Sigma'_{\phi'},\overline{\Sigma '}_{\phi'})),\bar F^{-1}(\bar X(\Sigma'_{\phi'},\overline{\Sigma'}_{\phi'}))\right)-\Sigma'_{\phi'} X(\Sigma'_{\phi'},\overline{\Sigma'}_{\phi'})-\overline{\Sigma'}_{\phi'}\bar X(\Sigma'_{\phi'},\overline{\Sigma'}_{\phi'})~,\\
&&
\eeqs
after solving
\be
K_X\frac {\partial F^{-1}}{\partial X}-\Sigma'_{\phi'}=0~, \quad K_{\bar X}\frac {\partial \bar F^{-1}}{\partial \bar X}-\overline{\Sigma}'_{\phi'}=0~.
\ee
Comparing to (\ref{first}) we see that $\Sigma'_{\phi'}$  as a functions of $X$ is related to $\Sigma_{\phi}$ as a function of $X$ via a holomorphic coordinate transformation depending on the K\"ahler gauge transformation $F$ and similarily for their complex conjugate. Explicitly
\beqs\nonumber
&&\Sigma_{\phi}=K_X(X,\bar X)\\[1mm]
&&\Sigma'_{\phi'}=K_X\left(F^{-1}(X), \bar F^{-1}(\bar X)\right)\frac {\partial F^{-1}(X)}{\partial X}~.
\eeqs
The relation between ${\cal U}(\Sigma_{\phi'},\overline{\Sigma'}_{\phi'})$ and $U(\Sigma_\phi,\overline{\Sigma}_\phi)$ is more complicated due to the linear $X$ terms, but can be worked out from (\ref{calU}) and (\ref{U}).


\end{document}